\documentclass[aps,prl,twocolumn,showpacs,superscriptaddress]{revtex4}
\usepackage[dvips]{graphicx}
\def\be{\begin{equation}}
\def\ee{\end{equation}}

\begin{document}
\title{Hysteresis loops and adiabatic Landau-Zener-St\"{u}ckelberg\\
transitions in the magnetic molecule $\{\textrm{V}_{6}\}$}
\author{I. Rousochatzakis}
\email{rusohatz@ameslab.gov}
\affiliation{Ames Laboratory and Department of Physics and Astronomy,
Iowa State University, Ames, Iowa, 50011}
\author{Y. Ajiro}
\affiliation{Department of Chemistry, Graduate School of Science, Kyoto University, Kyoto 606-8502, Japan}
\affiliation{
CREST, Japan Science and Technology Agency, Saitama 332-0012, Japan}
\author{H. Mitamura}
\affiliation{Institute for Solid State Physics, University of Tokyo, Chiba 106, Japan}
\author{P. K\"{o}gerler}
\affiliation{Ames Laboratory and Department of Physics and Astronomy,
Iowa State University, Ames, Iowa, 50011}
\author{M. Luban}  
\affiliation{Ames Laboratory and Department of Physics and Astronomy,
Iowa State University, Ames, Iowa, 50011}

\begin{abstract}
We have observed hysteresis loops and abrupt magnetization steps in the magnetic molecule
$\{\textrm{V}_{6}\}$, where each molecule comprises a pair of identical spin triangles, 
in the temperature range 1-5 K for external magnetic fields $B$ with sweep rates of several 
Tesla/ms executing a variety of closed cycles. The hysteresis loops are accurately reproduced using a 
generalization of the Bloch equation based on direct one-phonon transitions 
between the instantaneous Zeeman-split levels of the ground state (an $S=1/2$ doublet) of each spin triangle.
The magnetization steps occur for $B\approx 0$ and they are explained in terms of adiabatic 
Landau-Zener-St\"{u}ckelberg transitions between the lowest magnetic energy levels 
as modified by inter-triangle anisotropic exchange of order 0.4 K.  
\end{abstract}

\pacs{75.50.Xx, 75.45.+j, 71.70.-d}
\maketitle
Magnetic molecules provide a very convenient platform for exploring fundamental issues
in nanomagnetism. Heisenberg exchange between the magnetic-ion spins embedded in each molecule
gives rise to a discrete spectrum of magnetic energy levels.
Moreover, the magnetic interaction (dipole-dipole) between molecules
is generally so small as compared to intra-molecular exchange interactions that a crystal sample
may be regarded as a macroscopic assembly of independent identical quantum nanomagnets.
One significant goal is to understand the interactions of the magnetic molecules
with the environment (``heat bath''), for example via phonons. In particular, it is essential 
to understand the nature of the thermal relaxation mechanism, the controlling factors
responsible for irreversible and dissipative phenomena, and the detailed route to thermal 
equilibrium of these nano-size quantum spin systems. The simpler the spin system the greater the 
prospects for achieving a deep understanding of the underlying issues, and this opportunity is 
provided by the magnetic molecule $\{\textrm{V}_{6}\}$\cite{Luban}. Each $\{\textrm{V}_{6}\}$ includes
a pair of triangles of exchange-coupled vanadyl (VO$^{2+}$, spin 1/2) ions. 
As shown below, at low temperatures the 
instantaneous magnetization, $M(t)$, of this spin system, in response to pulsed magnetic 
fields, $B(t)$, with sweep rates of several Tesla/ms, exhibits pronounced hysteresis loops as 
well as abrupt magnetization steps that are due to Landau-Zener-St\"{u}ckelberg (LZS) 
transitions\cite{LZS,Miyashita-LZS} between lowest energy levels. 
By explaining the details of the dynamical magnetization
one establishes both the low-temperature relaxation mechanism for the individual magnetic molecules 
as well as microscopic information concerning the lowest energy levels, not readily accessible. 
Indeed, our analysis suggests 
the existence in this magnetic molecule of an effective inter-triangle anisotropic exchange of order 0.4 K;
otherwise Kramers' theorem\cite{Abragam,Miyashita-Kramers} would forbid the occurrence of 
LZS transitions.

There are several important differences between the present work and previous studies of $M(t)$ 
in magnetic molecules in time-dependent magnetic fields. From our observation of hysteresis effects in 
$\{\textrm{V}_{6}\}$ we conclude that the thermal relaxation time $\tau$ in this molecule is of order 0.1 ms.
This is many orders of magnitude shorter than those reported for ``single-molecule magnets'' 
such as $\{\textrm{Mn}_{12}\}$\cite{Sessoli} and $\{\textrm{Fe}_{8}\}$\cite{Sangregorio} where a large 
anisotropy energy barrier is responsible for relaxation times of order $10^{3}-10^{5}$ sec. 
Also, we assume that the phonon bottleneck effect which typically occurs at low temperatures 
(e.g. $T<200$ mK for $\{\textrm{V}_{15}\textrm{As}_{6}\}$\cite{Chiorescu}) does not arise:  
For the temperatures of our experiment ($T>1.5$ K) the number of available resonant phonons 
per molecule is large so that they equilibrate independently from the spins (typical 
times $\tau_{ph} < 10^{-6}$ s, much smaller than both the experimental time scale $\tau_{exp}\sim 1$ ms and
the relaxation times $\tau$ of the spins). Moreover, due to the high sweep rate of $B(t)$ 
in our measurements, LZS transitions are consequential 
only in the immediate vicinity of $B=0$. Away from $B=0$, we use a generalization of the standard
Bloch equation for $M(t)$, where the relaxation rate depends on the instantaneous $B(t)$. 
The excellent agreement obtained between theory and experiment allows us to identify 
the dominant mechanism for thermal relaxation in terms of direct one-phonon processes. 
To our knowledge, this is the first time that 
quantitative agreement between theory and experiment has been achieved for hysteresis loops 
in magnetic molecules. 

We first summarise the most important known features of $\{\textrm{V}_{6}\}$\cite{Luban}. 
The magnetic molecule [H$_{4}$V$_{6}^{\textrm{IV}}$O$_{8}$(PO$_{4}$)$_{4}$
\{(OCH$_{2}$)$_{3}$CCH$_{2}$OH\}$_{2}$]$^{6-}$,
abbreviated as $\{\textrm{V}_{6}\}$, and isolated as 
(CN$_{3}$H$_{6}$)$_{4}$Na$_{2}\{\textrm{V}_{6}\}\bullet 14$ H$_{2}$O,  
may be pictured (see Fig. 1) in terms of two identical triangular 
units per molecule, each unit consisting of three spins $s=1/2$ (VO$^{2+}$ ions) interacting via isotropic 
antiferromagnetic (AFM) exchange. 
Two of the 2-spin exchange constants (shown in blue) are equal ($J_{a}\approx 65$ K in units of $k_{B}$), 
and an order of magnitude larger than the third (shown in red, $J_{c}\approx 7$ K). 
Additionally, from nuclear magnetic resonance (NMR) studies and chemical structure analysis it has been 
argued that there exists a very weak inter-triangle
exchange interaction (yellow bonds, approximately 0.3 K). 
In the absence of inter-triangle exchange,
for $B=0$ the ground state of each triangle consists of a 2-fold degenerate doublet with 
total spin $S=1/2$, consistent with Kramers' theorem. The excited levels are a second degenerate doublet 
with $S=1/2$ and excitation energy $(J_{a}-J_{c})\approx 58$ K, and a 4-fold degenerate level with 
 $S=3/2$ and excitation energy (also measured from the ground state) $3J_{a}/2 \approx 97$ K. 
\begin{figure}[!t]
\centering
\includegraphics[width=3.8in, height=2.5in]{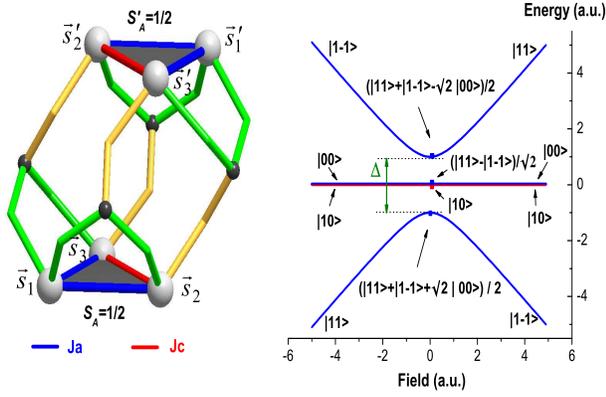}
\caption{Left: Structure of the two spin triangles in the  
$\{\textrm{V}_6\}$ anion (bright grey spheres). Phosphate exchange 
paths (green), and other ligands not shown, mediate strong intra-triangle exchange ($J_{a}$, blue) and weak 
inter-triangle exchange (yellow bonds). At low temperatures each triangle
behaves as a spin 1/2 entity ($S_A = S'_{A} =1/2$). 
Right: Energy diagram for one scenario of inter-triangle exchange where a term of the form 
$(\Delta/2)(S_{Ax} S'_{Az}-S_{Az} S'_{Ax})$ admixes several of the states $|S, M_S; S_A=1/2, S'_A=1/2>$,
shown as blue lines.} 
\end{figure}
In the experiments described below we consider 
temperatures in the range 1.5-5 K and $B<25$ Tesla, well below the field value ($\approx 74$ Tesla) when the 
$S=3/2, M_{S}=-3/2$ level crosses the ground state $S=1/2, M_{S}=-1/2$ level.
As such, it suffices to consider only the ground state doublet of each triangular unit. 
A weak residual inter-triangle anisotropic exchange will lift the 4-fold degeneracy for $B=0$ of
each molecule and give rise (in general) to four distinct energy levels (see below). As remarked above,
the occurrence of these splittings can be manifest when the molecules are subject to pulsed 
magnetic fields, giving rise to a sudden reversal in magnetization when the field crosses $B=0$, 
as a result of LZS transitions between the split levels. Apart from the vicinity of $B=0$ 
the magnetic properties at low $T$ of a $\{\textrm{V}_{6}\}$ sample may be accurately described in terms of an 
ensemble of independent $S=1/2$ spin triangles.

Time-resolved magnetization measurements were performed on a powdered sample for half-cycle and 
full-cycle sweeps by a standard induction method using compensated pickup coils and a nondestructive 
long pulse magnet installed at ISSP. Utilizing fast digitizers, the inductive method provides data for 
$dM/dt$ and $dB/dt$ which are subsequently integrated to give results for $M$ versus $B$. The pulsed fields 
have a nearly sinusoidal shape as a function of time, with a half-period about 21 ms (0-maximum-0). 
The sample of 36.9 mg was packed in a thin-walled cylindrical teflon capsule (inner diameter 3.0 mm) 
and then directly immersed in a liquid Helium bath.
\begin{figure}[!h]
\centering
\includegraphics[width=3.5in, height=5in]{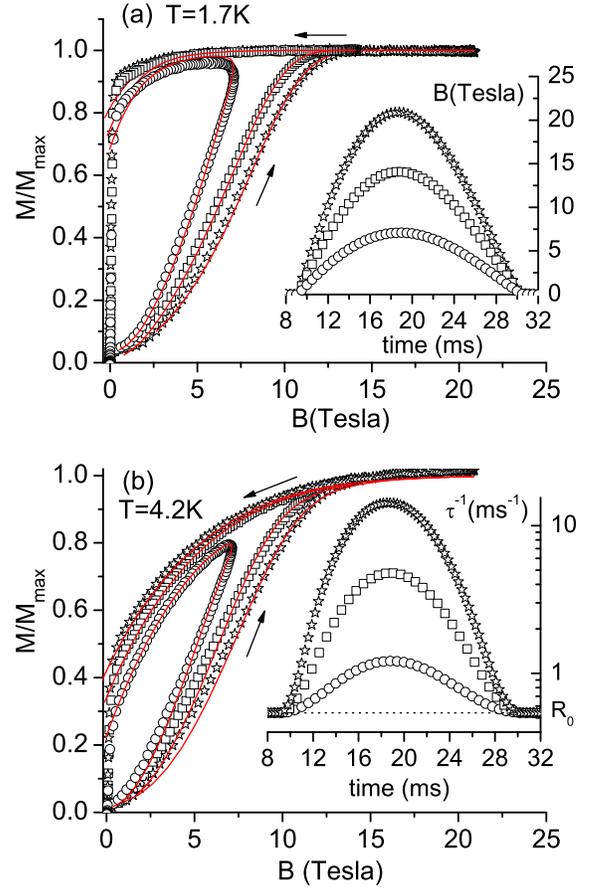}
\caption{Measured magnetization vs magnetic field for $T=1.7$ K (a) and $T=4.2$ K (b), for the three
half-cycle sweeps shown in the inset of Fig. 2(a).
The solid red lines are obtained using Eqs. (1) and (2). 
The time dependence of the relaxation rate $\tau^{-1}$ according to Eq. (2) for $T=4.2$ K is given in the 
inset of Fig. 2(b). The dotted line indicates the residual constant $R_{0}$ in Eq. (2). 
The sudden drop of $M$ to zero at $B=0$ is explained in the text in the context of LZS transitions.}
\end{figure}

In Figs. 2 and 3 we present our experimental and theoretical results for the magnetization versus applied magnetic
field for two different temperatures (1.7 K and 4.2 K) and for half-cycle and full-cycle sweeps shown
in the insets of Fig. 2(a) and Figs. 3(a) and 3(b). The two striking features of the $M$ vs. $B$ data
are hysteresis loops and the appearance of magnetization steps (in Fig. 2) and near-reversals
(Fig. 3) in the immediate vicinity of $B=0$. The hysteresis loops (all data except in the immediate vicinity 
of $B=0$) are reproduced (solid lines in Figs. 2 and 3) by numerical solution of the following 
generalization\cite{Rousochatzakis} of the familiar Bloch equation\cite{Bloch,Abragam}
\be
\frac{d}{dt}M(t)=\frac{1}{\tau(T,B(t))}[M_{eq}(T,B(t))-M(t)],
\ee
with the relaxation rate $1/\tau$ given by
\be
\frac{1}{\tau(T,B(t))}=\frac{3 (g \mu_{B})^3 V_{sl}^2}{2\pi\rho v^5\hbar^4}B(t)^3\textrm{coth}[\frac{g\mu_{B}B(t)}{2k_{B}T}] + R_{0}.
\ee
Here $\mu_{B}$ is the Bohr magneton, $\rho$ denotes the mass density, 
$v$ the sound velocity, $V_{sl}$ the characteristic modulation of
the spin energy under long-wavelength acoustic deformation, and $M_{eq}(T,B(t))$ is the standard 
two-level equilibrium magnetization for an instantaneous field $B(t)$ and for temperature $T$, \emph{i.e.}, 
$M_{eq}(T,B(t))/M_{max}= \textrm{tanh}[g\mu_{B}B(t)/(2k_{B}T)]$, 
where $M_{max}= 2(N_{A}g\mu_{B}/2)$. 
We have derived Eq. (1) from first principles\cite{Rousochatzakis} 
upon making the assumption that for these temperatures the phonons are in thermal equilibrium with the cryostat at all 
experimental times. 
The first term of Eq. (2) is the low-temperature relaxation rate of the spins due to direct one-phonon processes,
where spin flips are triggered by an acoustic phonon mode meeting the resonance condition
for the \textit{instantaneous} 
energy separation of the two-level spin system.
This term is a generalization of the standard expression for the relaxation rate due to one-phonon 
processes in a static external field\cite{Abragam}. Both the $B(t)^{3}$ factor, proportional 
to the phonon energy density, and the statistical mechanical factor
depend on the instantaneous resonance frequency, proportional to $B(t)$. 
The numerical value of $V_{sl}$ depends on the specific details of the spin-phonon coupling 
(see, for example, discussion for paramagnetic spins in Ref. \cite{Abragam}), which at present is unclear. 
The quantity $R_{0}$ in Eq. (2) represents additional 
relaxation processes present and it is taken as a fitting parameter. 
Using the measured value of $\rho=1.93$ g/cm$^3$ and estimating $v=3000$ m/s, we obtain excellent 
agreement with our data for the choices
$R_{0}=0.2$ ms$^{-1}$ for $T=1.7$ K and $R_{0}=0.5$ ms$^{-1}$ for
$T=4.2$ K and $V_{sl}/k_{B}= 0.35$ K. 
Despite the smallness of $R_{0}$ it is important to retain this term in order to 
achieve a good fit to the experimental data in the low-field regime (below 4 Tesla for 1.7 K and below
7 Tesla for $T=4.2$ K); for higher fields the dominant contribution to $1/\tau$ comes from the one-phonon
term. We emphasize that the solution of Eq. (1) is extremely sensitive to the explicit functional form of the first 
term of Eq. (2)): Adopting a different choice of functional form one cannot achieve quantitative agreement 
with the observed hysteresis loops, for the whole field range and for different choices of field sweeps. 
Achieving an excellent fit to our measured data for a variety 
of choices of $B(t)$ thus affirms the basic correctness of these two equations.   
\begin{figure}[!t]
\centering
\includegraphics[width=3.5in, height=5in] {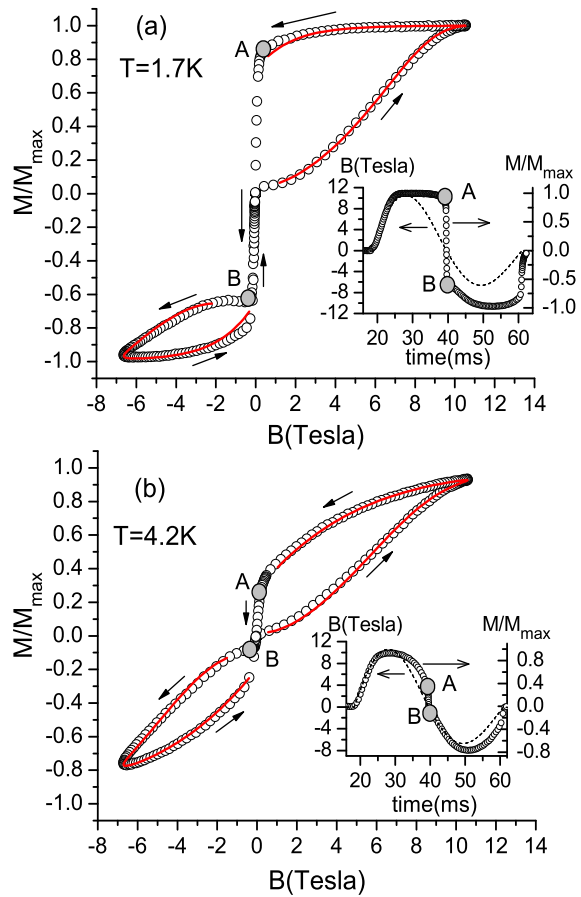}
\caption{Measured magnetization vs magnetic field for $T=1.7$ K (a) and $T=4.2$ K (b), 
for the full-cycle sweep shown in the inset of Figs. 3(a) and 3(b) (dotted lines).  
The solid lines are obtained using Eqs. (1) and (2). The insets also show the 
measured magnetization (circles) vs time. The LZS transitions
occur in the interval between the points A and B.}
\end{figure}

We now discuss the magnetization steps observed for $B\approx 0$, 
and in particular the interval between points A and B in Figs. 3(a) and 3(b)
 (the other steps seen in Figs. 2 and 3 have the same physical origin and will not be discussed separately). 
 In this interval the external field varies approximately linearly with time, with sweep rates of order 
1 Tesla/ms. We note that at point A, $M_{A}/M_{max}$ is somewhat less than unity due to thermal relaxation 
before entering the fast-reversal regime. Equivalently, at point A we are dealing with a statistical mixture
of spin-up and spin-down states.
The most striking feature though is that the magnetization $M_{B}$, at point B, nearly equals   
$-M_{A}$. 
We find that the time-widths of the near-reversals is shorter the faster the sweep rate, and is in the range 
0.5-0.8 ms. We propose that adiabatic LZS transitions are responsible for the magnetization steps observed 
in our system. The characteristic energy gap $\Delta$ of the LZS 2-level\footnote{We anticipate that a similar 
relation holds for the general 4-level LZS problem.} model is related to the time-width
$\delta t_{LZS}$ of the magnetization step and the field sweep rate $r$ by the relation 
$\delta t_{LZS}=2\Delta /(g\mu_{B} r)$ \cite{LZS,Miyashita-LZS}).
Thus, the measured time-widths of the steps give, as a first estimate for the zero-field energy gap, 
$\Delta\approx 0.4$ K. Using the above estimate for $\Delta$, 
we are indeed in the regime of adiabatic LZS transitions, since the transition probability
$P_{LZS}=1-exp(-\pi (\Delta^{2}/(2\hbar g \mu_{B} r)) \approx 1$ \cite{LZS,Miyashita-LZS}, 
thus implying that $M_{B}=-M_{A}$. The observed deviation of $M/M_{max}$ from exact reversal is about $15\%$
 for $T=1.7$ K. This discrepancy may be due to the role of the heat bath, i.e., 
the problem of dissipative LZS transitions (see, for example, \cite{Nishino} and references therein); however, a
systematic investigation of this issue is in progress. 
More generally, it should be noted that, according to the above formula for $\delta t_{LZS}$, 
it is the high sweep rates used in our experiment that ensure that the adiabatic LZS transitions take place over 
such a short time interval as to be clearly distinct from the hysteresis loops of the thermal relaxation regime. 
 
The origin of LZS transitions in $\{\textrm{V}_{6}\}$ remains to be discussed. 
The relatively large estimated value (0.4 K) of the zero-field energy gap $\Delta$
for these molecules suggests that its origin cannot be due to dipolar or 
hyperfine fields. In addition, as explained above, the lowest energy levels of each \textit{independent} triangle 
are doubly degenerate for $B=0$ (and not four-fold degenerate as in 
$\{\textrm{V}_{15}\textrm{As}_{6}\}$\cite{Chiorescu}) consistent with Kramers' theorem. 
Hence, for $B=0$ the ground state of such a molecule 
would consist of four degenerate states, namely the three symmetric states of the $S=1$ 
triplet and the antisymmetric $S=0$ singlet state.
Inter-triangle exchange coupling could lift this degeneracy and give rise to avoided level crossings. 
However, since isotropic inter-triangle exchange cannot account for admixing states of total spin, 
we suggest that the avoided level crossings are due to the anisotropic (symmetric 
or antisymmetric) portion. One scenario is given in Fig. 1. The behavior of the dynamical magnetization 
thus involves LZS transitions between (at most) four levels. A detailed theoretical treatment
of these transitions will be given elsewhere. 

In summary, time-resolved magnetization measurements using sweep rates of order 1 Tesla/ms
show hysteresis loops and magnetization steps for $B\approx 0$ in the magnetic molecule $\{\textrm{V}_{6}\}$. 
The two effects are clearly distinct because of the relatively high sweep rates used in our experiment.
In the absence of both an anisotropy energy barrier and the phonon bottleneck effect, 
the hysteresis effects exhibited by this molecule
occur because the spin relaxation times are of the experimental time scale.    
Using a generalization of the Bloch equation we were able to reproduce our experimental data for 
$T=1.7$ K and $T=4.2$ K for a large variety of field sweeps, and thus identify direct
one-phonon resonant transitions among the Zeeman-split doublet of each triangle 
as the dominant mechanism underlying the hysteresis behavior. The main assumption of our model, namely 
that the phonons are in equilibrium with the cryostat, should break down for temperatures
below 1 K due to the phonon bottleneck effect. 
In fact, our preliminary data at $T=0.6$ K indicate that the relaxation rate $1/\tau$ deviates from Eq. (2). 
A systematic investigation of this issue is in progress.  
The steps of the magnetization for $B\approx 0$ are attributed to adiabatic LZS transitions between lowest magnetic 
energy levels impacted by the existence of anisotropic inter-triangle exchange interaction of order 0.4 K. 
This estimate is consistent with that previously suggested by NMR data\cite{Luban}. 
A more precise value of $\Delta$ could possibly be determined by specific heat 
measurements\cite{Affronte}, or by Electron Paramagnetic Resonance (EPR)
techniques. The small departures from complete magnetization reversal suggests that one cannot entirely neglect 
the role of the heat bath. 
More generally, exploring nanomagnets with pulsed magnetic fields can reveal a variety of 
fascinating dynamical phenomena and provide microscopic information that otherwise is not readily accessible.      
  
Ames Laboratory is operated for the U.S. Department of Energy by Iowa State University
under Contract No. W-7405-Eng-82.

\end{document}